\documentclass[conference, a4paper]{IEEEtran}
\IEEEoverridecommandlockouts
\usepackage{color}
\usepackage{tikz}
\usepackage{pgfplots}
\pgfplotsset{compat=1.12}
\usetikzlibrary{patterns}
\usepackage{pgfplotstable}
\usepackage{csvsimple}
\definecolor{excelblue}{RGB}{94,156,211}
\definecolor{excelorange}{RGB}{235,125,60}
\definecolor{excelgray}{RGB}{165,165,165}
\definecolor{excelgreen}{RGB}{114,172,77}

\usepackage{cite}
\usepackage{amsmath,amssymb,amsfonts}
\usepackage{algorithmic}
\usepackage{graphicx}
\usepackage{textcomp}
\usepackage{xcolor}
\usepackage{optidef}
\usepackage{subcaption}
\def\BibTeX{{\rm B\kern-.05em{\sc i\kern-.025em b}\kern-.08em
    T\kern-.1667em\lower.7ex\hbox{E}\kern-.125emX}}

\makeatletter
\newcommand{\linebreakand}{%
  \end{@IEEEauthorhalign}
  \hfill\mbox{}\par
  \mbox{}\hfill\begin{@IEEEauthorhalign}
}
\makeatother
\begin{document}

\title{AI Driven Near Real-time Locational Marginal Pricing Method: A Feasibility and Robustness Study\\
{}
\thanks{}
}

\author{\IEEEauthorblockN{Naga Venkata Sai Jitin Jami}
\IEEEauthorblockA{\textit{Institute of Computing (CI)} \\
\textit{Università della Svizzera italiana}\\
Lugano, Switzerland\\
jitin.jami@usi.ch}
\and
\IEEEauthorblockN{Juraj Kardoš}
\IEEEauthorblockA{\textit{Institute of Computing (CI)} \\
\textit{Università della Svizzera italiana}\\
Lugano, Switzerland \\
juraj.kardos@usi.ch}
\and
\IEEEauthorblockN{Olaf Schenk}
\IEEEauthorblockA{\textit{Institute of Computing (CI)} \\
\textit{Università della Svizzera italiana}\\
Lugano, Switzerland \\
olaf.schenk@usi.ch}
\linebreakand
\IEEEauthorblockN{Harald Köstler}
\IEEEauthorblockA{\textit{Department of Computer Science} \\
\textit{Friedrich-Alexander-Universität}\\
Erlangen, Germany \\
harald.koestler@fau.de}
}

\maketitle

\begin{abstract}
%Locational Marginal Pricing (LMP) is a pricing mechanism used to calculate the cost of providing electricity to a specific point in the electricity grid, and accurate LMP predictions are essential for market participants to optimize their operation schedules and bidding strategies.
Accurate price predictions are essential for market participants in order to optimize their operational schedules and bidding strategies, especially in the current context where electricity prices become more volatile and less predictable using classical approaches. 
The Locational Marginal Pricing (LMP) pricing mechanism is used in many modern power markets, where the traditional approach utilizes optimal power flow (OPF) solvers. However, for large electricity grids this process becomes prohibitively time-consuming and computationally intensive. Machine learning (ML) based predictions could provide an efficient tool for LMP prediction, especially in  energy markets  with intermittent sources like renewable energy.
%Therefore, there has been an effort in recent years to explore faster prediction of LMPs using machine learning techniques. 
This study evaluates the performance of popular machine learning and deep learning models in predicting LMP on multiple electricity grids. The accuracy and robustness of these models in predicting LMP is assessed considering multiple scenarios. The results show that ML models can predict LMP 4-5 orders of magnitude faster than traditional OPF solvers with 5-6\% error rate, highlighting the potential of ML models in LMP prediction for large-scale power models with the assistance of hardware infrastructure  like multi-core CPUs and GPUs in modern HPC clusters.
%with no topology changes. Moreover, they tend to have similar error rates on edge-case scenarios with topology changes. 
%The study also demonstrates promising signs of robustness when neural networks did not have a drop in performance against edge-case scenarios on larger electricity grids.
\end{abstract}


\section{Introduction}
Nodal Pricing, also known as Locational Marginal Pricing (LMP), is a widely-used pricing mechanism in energy markets around the world, including the US, Ireland, New Zealand, and Singapore \cite{LMPBook}. It calculates nodal level remuneration based on location, acknowledging the importance of location when it comes to electricity price \cite{Cantillo2022}. Accurate forecasting of LMPs is, therefore, essential for market participants, such as balancing and flexibility service providers, to optimize the scheduled operation and bidding strategy \cite{Zheng2020}. A regular power system is generally very large and complex, thus computing the LMP of each node becomes prohibitively expensive.  To address this challenge, machine learning tools can be leveraged not only to predict prices, but also to guide their operational schedules based on changing market conditions.% by training them on real-time data from sensors and other sources.

The increasing penetration of renewable energy sources in the energy mix, such as solar and wind, increases the volatility and unpredictability of electricity prices. ML models can help to mitigate these uncertainties and enable stakeholders in the energy market to make more informed decisions. The LMP is usually determined by solving the direct current optimal power flow (DC-OPF) problem, a simplified version of optimal power flow (OPF) which is a constrained nonlinear programming problem \cite{bai2019review}. The large-scale nature of OPF problems exhibits excessive hardware and computational-time requirements, especially when dealing with real-time predictions of realistic electricity grids \cite{Zimmerman2011}. In order to retain the operational standards of the power grid operators or the market participants, they need to re-run the OPF very often, following the forecast updates of the renewable energy sources.

Price prediction and price forecasting tasks are increasingly relying on ML methodologies to reduce processing time \cite{Cantillo2022}. Solution of ML models can also be used as warm-start points for solution using iterative solvers. Models can be tailored towards solving for a subset of the OPF solution like system voltages, generator scheduling or LMP prediction. With our proposed ML model solution, the ML model can be pre-trained on the data samples around the most-likely scenarios based on grid-specific data and the on-line prediction for any updated scenario can be obtained from the pre-trained ML model.

In previous studies, Graph Neural Networks were used in \cite{Liu2022} to create a topology-aware Neural Network (NN) so that any changes to the grid do not trigger retraining of the model or hamper prediction capabilities due to over-dependence on historical data. In \cite{baker2019learning}, the author tries to solve the AC-OPF problem with random forests to predict voltage and generation solutions that would serve as an intelligent, warm start for further solving the AC-OPF. In \cite{zamzam2020learning}, the authors use a NN to predict optimal AC-OPF solution for 3 IEEE standard electricity grids. They also comment on the feasibility of the solution and the amount of time saved. The authors of \cite{zhao2021ensuring} focuses further on implementing hard feasibility strategies to ensure the predicted solutions are within the feasible space. Their subsequent paper \cite{pan2021deepopf} uses similar strategies on the DC-OPF formulation.  DC-OPF formulation was directly solved using a convex NN in \cite{zhang2021convex} and used the KKT conditions to construct a loss metric for training. %A Sensitivity Informed DNN was implemented to match the OPF optimisers and the partial derivatives concerning OPF parameters.
In \cite{Garcia2022}, the authors use a blend of seasonal, weather, economic and operational data to predict LMP value in Mexico using ANNs and global senstivity analysis (GSA) to determine the importance of external factors. Transformers, a well-known sequence to sequence(seq2seq) architecture for deep learning, were used to predict LMP in the PJM power market using historical data. 

%The authors in their papers \cite{venzke2020learning}, \cite{nellikkath2021physics} and \cite{nellikkath2021dcphysics} use a Physics Informed DNN that works in 2 stages, one stage to introduce KKT conditions of the OPF into the DNN and another stage of extracting worst-case guarantees for generation and line flow constraint violations. %In \cite{Cantillo2022}, the authors use Support Vector Regression to predict LMP over data generated by Monte-Carlo simulations of an example case on the MATPOWER library.

Accuracy of several popular ML and deep learning models proposed in the literature is studied in context of LMP price prediction based on data involving demand, supply and generator cost curves. Various edge-case scenarios and the grid topology adjustments are considered to test model robustness and assess the performance scalability to larger electricity grids. 
The ML models considered in this study are Decision Tree Regression (DTR), Gradient Boosting Regression (GBR), Random Forest Regression (RFR), and Deep Neural Network with multiple hidden layers (NN - 1 \& 2). 
When assessing the accuracy and robustness of different ML models, there are a few key factors to consider. One important factor is the model's predictive performance, which can be measured using metrics like Mean Absolute Percentage Error (MAPE). It is also essential to consider the robustness of the model, which refers to its ability to perform well on a wide range of input data. This can be evaluated by testing the model on various data sets that simulate changes in market conditions and other external factors. The scalability of the accuracy and robustness can be tested by running experiments on electricity grids of various sizes.

The remaining part of this paper is structured as follows. Sec.~\ref{sec:dg} explains the data generation methodology used to generate training and testing data. Sec.~\ref{sec:ne} gives an overview on the numerical experiments conducted and the results of the experiments. Finally, Sec.~\ref{sec:con} provides concluding remarks and directions for future work.

\section{Data Generation} \label{sec:dg}

All the training and testing ground truth data is generated with the help of MATLAB. The electricity grids of choice are chosen from PGLib-OPF \cite{babaeinejadsarookolaee2019power}. The PGLib-OPF models contain static data, i.e., a snapshot of the grid state in a single time instance, consisting of individual voltage levels across nodes, grid topology, power injection and power withdrawal at nodes. This single snapshot translates into a single DC-OPF optimization problem imported into MATPOWER \cite{Zimmerman2011} that is solved using an optimization solver such as MOSEK \cite{MOSEK}. The choice of the electricity grid and their properties are detailed in Tab. \ref{tab:2.1}.
%\begin{enumerate}
%    \item \textbf{case30}: IEEE power flow data for 30 bus, 6 generator case.
%    \item \textbf{case240}: PSERC M-21 power flow data for 240 bus, 143 generator case.
%    \item \textbf{case1354}: PEGASE - Pan European Grid Advanced Simulation and State Estimation data.
%    \item \textbf{case1888}: AC Power flow data for French system.
%\end{enumerate}

\begin{table}[t!]
\centering
\caption{Description of electricity grids under consideration.}
    \begin{tabular}{l|r|r|r}
        \hline
         \textbf{Case} & \textbf{\# load buses} & \textbf{\# generator buses} & \textbf{\# branches}\\
         \hline
         \hline
         \texttt{case30} & 30 & 6 & 41\\
         \hline
         \texttt{case240} & 240 & 143 & 448\\
         \hline
         \texttt{case1354} & 1354 & 260 & 1991\\
         \hline
         \texttt{case1888} & 1888 & 296 & 2531\\
         \hline
    \end{tabular}
    \label{tab:2.1}
\end{table}

The PGLib-OPF provides a single feasible snapshot for each electricity grid. However, training a ML model will require more than a single snapshot (or single instance) that have to be dissimilar from each other. In order to generate these heterogeneous instances of data, perturbations at the nodal level are introduced. When the active loads are perturbed at all nodes,  the resulting DC-OPF problem differs from the base case also in terms of the nodal prices. The nodal level perturbations $s_{P_d}$ are calculated using the following equation:
\begin{equation} \label{eq:nodal_perturb}
    s_{P_d} = 1 + \frac{s_{grid} \times s_{nodal}}{100},
\end{equation}
where the scalar $s_{grid}$ is the nodal perturbation expressed in percent relative to the base case and is constant for all the grid nodes in a given problem instance. The scaling factor $s_{nodal}$ is a vector of uniformly distributed random numbers in the range (0.9, 1.1) with the length equal to the number of the grid nodes.
The motivation behind the nodal perturbation Eq.~\eqref{eq:nodal_perturb} is to simulate a given global trend ($s_{grid}$) in the grid, such as an increase or decrease of the active load, but this is not equal across all the nodes in the grid. Thus, we apply random noise at the nodal level ($s_{nodal}$) to these global trends to achieve the final scaling factor.

\begin{table}[th!]
\centering
    \caption{Range of the global perturbations $s_{grid}$.}
    \begin{tabular}{l|r}
        \hline
         \textbf{Case} & \multicolumn{1}{l}{\textbf{Perturbation Range (\%)}}\\
         \hline
         \hline
         \texttt{case30} & \{-30, 30\}\\
         \texttt{case240} & \{-70, -10\}\\
         \texttt{case1354} & \{-50, 0\}\\
         \texttt{case1888} & \{-40, 10\}\\
         \hline
    \end{tabular}
    \label{tab:4.3}
\end{table}

\subsection{Training Data}
The base snapshot of each grid is first manipulated with the help of Eq.~\eqref{eq:nodal_perturb} to generate up to 5000 heterogeneous instances. The $s_{grid}$ parameter was randomly sampled from the given range for each problem instance.
Each instance corresponds to a separate optimization problem solved with MATPOWER and MOSEK. We extract the information to use as input features $x_i$ for training the ML model. The input features of each node (demand and generator nodes) on the electricity grid are active power demands $P_d$  at each load bus and transmission capacity factor $P_l$ is defined as
%\begin{itemize}
%    \item active power demands at each load bus,
%    \item transmission capacity factor,
%\end{itemize}
\begin{equation}
        P_l= \frac{P_d}{P_l^{max}},
\end{equation}
where $P_d$ represents nodal active power demand and $P_l^{max}$ represents the maximum capacity of the connected transmission lines of a given node.
The input features are not part of the solution of MATPOWER and MOSEK, they are part of the problem definition.

\subsection{Testing Data} \label{sec:test_data}
The generation of test data is crucial to our objectives of measuring accuracy and testing robustness. We achieve this by generating 4 testing datasets described in Tab ~\ref{tab:4.4}. Data for each case was generated the same way our training data was but for only 100 heterogeneous instances. We also choose the same input features. However, the data for each instance in this dataset has been generated by further manipulating the electricity grid apart from introducing nodal perturbations.

The test cases aim to test the robustness of trained ML models under various edge-case scenarios. All the data has been generated by first introducing the same grid level perturbations shown in Tab.~\ref{tab:4.3}. The test case 1 has no further manipulation. It has the same characteristics as training data but the specific instances have not been seen by the models during the training case.
The remaining cases have been chosen to mimic real life scenarios like maintenance cycles or outages of the power grid components and simulate additional edge-case scenarios commonly occurring in security analysis of the power grid in the literature. 

\begin{table}[t!]
\centering
\caption{Test cases.}
\resizebox{0.45\textwidth}{!}{
    \begin{tabular}{l|l}
        \hline
         \textbf{Test Case} & \textbf{Description}\\
         \hline
         \hline
         \texttt{1} & Base case\\
         \hline
         \texttt{2} & 10\% reduction on transmission capacity\\
         \hline
         \texttt{3} & Missing 1 transmission line at random\\
         \hline
         \texttt{4} & Missing 1 generator at random\\
         \hline
    \end{tabular}}
    \label{tab:4.4}
\end{table}

\section{Numerical Experiments} \label{sec:ne}
The simulations are performed on the ICS cluster at USI, Lugano, which consists of 41 nodes equipped with two 10-core Intel Xeon E5-2650 v3 with frequency 2.30GHz. The nodes have 128 GB RAM memory. The language and library stack used for this projects are: Python 3.7, MATLAB R2020a, MATPOWER 7.1, MOSEK 10, PyTorch 10.1, Scikit-Learn 1.0.2, Bayesian Optimization 1.4.2 and Scikit-Optimize 0.9.0.
Training and testing data was generated using MATPOWER and MOSEK. Hyper-parameters tuning for all our ML models were conducted using bayesian optimisation using a subset of the training data to decide the hyper-parameters used for the experiments. The hyper-parameters of all the models are shown in Tab. \ref{tab:dtr2}.

\begin{table}[!t]
\caption{Hyper-parameters for ML models}
\centering
\resizebox{0.45\textwidth}{!}{
    \begin{tabular}{l|r|r|r}
        \hline
         \textbf{DTR} & \multicolumn{1}{l|}{\textbf{max\_leaf\_nodes}} & \multicolumn{1}{l|}{\textbf{min\_samples\_leaf}} & \multicolumn{1}{l}{\textbf{min\_samples\_split}}\\
         \hline
         \hline
         \texttt{case30} & 110 & 130 & 120\\
         \hline
         \texttt{case240} & 60 & 190 & 170\\
         \hline
         \texttt{case1354} & 60 & 190 & 170\\
         \hline
         \texttt{case1888} & 110 & 130 & 120\\
         \hline
    \end{tabular}}
    \label{tab:dtr2}
\vspace{0.2cm}
\resizebox{0.45\textwidth}{!}{
    \begin{tabular}{l|r|r|r|r}
        \hline
         \textbf{RFR} & \multicolumn{1}{l|}{\textbf{max\_leaf\_nodes}} & \multicolumn{1}{l|}{\textbf{min\_samples\_leaf}} & \multicolumn{1}{l|}{\textbf{min\_samples\_split}} & \multicolumn{1}{l}{\textbf{n\_estimators}}\\
         \hline
         \hline
         \texttt{case30} & 100 & 100 & 140 & 100\\
         \hline
         \texttt{case240} & 170 & 180 & 180 & 700\\
         \hline
         \texttt{case1354} & 30 & 190 & 70 & 300\\
         \hline
         \texttt{case1888} & 60 & 190 & 170 & 1200\\
         \hline
    \end{tabular}}
    \label{tab:rfr2}
\vspace{0.2cm}
\resizebox{0.45\textwidth}{!}{
    \begin{tabular}{l|r|r|r|r}
        \hline
         \textbf{GBR} & \multicolumn{1}{l|}{\textbf{learning\_rate}} & \multicolumn{1}{l|}{\textbf{max\_depth}} & \multicolumn{1}{l|}{\textbf{n\_estimators}} & \multicolumn{1}{l}{\textbf{subsample}}\\
         \hline
         \hline
         \texttt{case30} & 0.09 & 2 & 1500 & 0.2\\
         \hline
         \texttt{case240} & 0.09 & 2 & 1600 & 0.1\\
         \hline
         \texttt{case1354} & 0.01 & 4 & 200 & 0.1\\
         \hline
         \texttt{case1888} & 0.08 & 2 & 1800 & 0.1\\
         \hline
    \end{tabular}}
    \label{tab:gbr2}
\vspace{0.2cm}
\resizebox{0.45\textwidth}{!}{
    \begin{tabular}{l|r|r|r|r}
        \hline
         \textbf{DNN-1} & \multicolumn{1}{l|}{\textbf{\# hidden layers}} & \multicolumn{1}{l|}{\textbf{Topology}} & \multicolumn{1}{l|}{\textbf{Learning Rate}} & \multicolumn{1}{l}{\textbf{Batch Size}}\\
         \hline
         \hline
         \texttt{case30} & 1 & [128] & 0.009 & 128\\
         \hline
         \texttt{case240} & 2 & [128,64] & 0.006 & 128\\
         \hline
         \texttt{case1354} & 2 & [128,64] & 0.004 & 128\\
         \hline
         \texttt{case1888} & 3 & [4096,512,32] & 0.001 & 128\\
         \hline
    \end{tabular}}
    \label{tab:dnn12}
\vspace{0.2cm}
\resizebox{0.45\textwidth}{!}{
    \begin{tabular}{l|r|r|r|r}
        \hline
         \textbf{DNN-2} & \multicolumn{1}{l|}{\textbf{\# hidden layers}} & \multicolumn{1}{l|}{\textbf{Topology}} & \multicolumn{1}{l|}{\textbf{Learning Rate}} & \multicolumn{1}{l}{\textbf{Batch Size}}\\
         \hline
         \hline
         \texttt{case30} & 2 & [512,32] & 0.008 & 32\\
         \hline
         \texttt{case240} & 3 & [128,32,32] & 0.005 & 128\\
         \hline
         \texttt{case1354} & 3 & [128,32,16] & 0.005 & 128\\
         \hline
         \texttt{case1888} & 4 & [4096, 2048, 512, 32] & 0.001 & 128\\
         \hline
    \end{tabular}}
    \label{tab:dnn22}
\end{table}

The generated data is first pre-processed such that all features have zero mean and unit variance across all the instances of that experiment. In each experiment, 100 instances of the ML models were tested on the same testing dataset. Each model is first trained. Next, the LMP value of the test cases is predicted and the MAPE shown in Eq.~\eqref{eq:MAPE} is calculated. The error of the prediction is calculated against the ground truth, which is the LMP value calculated by solving the DC-OPF with MATPOWER and MOSEK. This metric expresses the average error of the predictions in the 100 trained model instances.

The MAPE is defined as
\begin{equation}
    MAPE(\Bar{y},\hat{y}) = \frac{100}{NS} \sum_{i=0}^{NS - 1} \frac{|\Bar{y}_i - \hat{y}_i|}{|\Bar{y}_i|},
    \label{eq:MAPE}
\end{equation}
where $NS$ is the number of samples used in the experiments, $\Bar{y}_i$ is the ground truth and $\hat{y}_i$ is the model prediction.

%\section{Results}
We focus on 3 different aspects of our study, namely processing time, training time and accuracy. Additionally, we aim to investigate the feasibility, robustness and scalability of using ML algorithms for the LMP predictions.

\subsection{Processing Time}
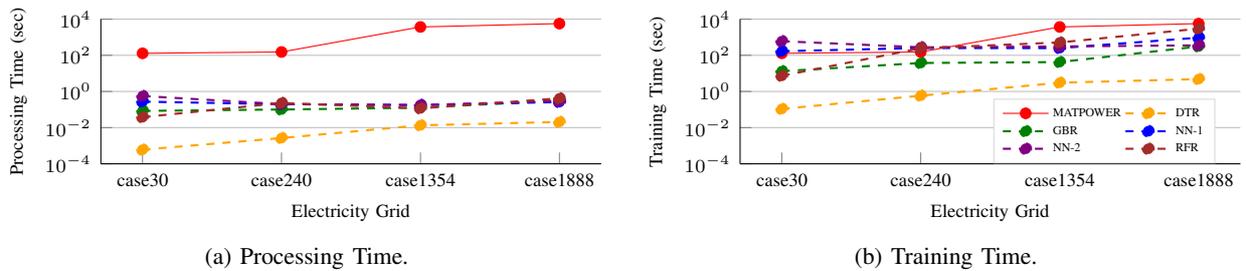
\begin{figure*}[t!]
	\centering
	\begin{subfigure}[t]{0.45\textwidth}
	    \centering
        \begin{tikzpicture}

% COLORS 
%4-class OrRd Tokyo SIAM PP18
\definecolor{mycolor0}{HTML}{FF0000}
\definecolor{mycolor1}{HTML}{FFA500}
\definecolor{mycolor2}{HTML}{008000}
\definecolor{mycolor3}{HTML}{0000FF}
\definecolor{mycolor4}{HTML}{800080}
\definecolor{mycolor5}{HTML}{A52A2A}

% \pgfplotstableread{figures/greasy/verification/dxt.dat}\data;

% \pgfplotstableread{figures/TcAvg/lpModelIntpnt/parsed_lp_intpnt_tc_avg.dat}\ipdata;

% \pgfplotstableread{figures/greasy/greasyLpIntpnt/avg_lp_intpnt_greasy_w_binding.dat}\lpintpntWbindingAvg;

\pgfplotstableread{grid time
1	128.5956
2	152.0848
3	3688.9000
4	5654.2000
}\MATLAB;

\pgfplotstableread{grid time
1	0.000605
2	0.002650
3	0.013370
4	0.020643
}\DTR;

\pgfplotstableread{grid time
1	0.084032
2	0.100437
3	0.126211
4	0.315168
}\GBR;

\pgfplotstableread{grid time
1	0.273739
2	0.197386
3	0.176523
4	0.259318
}\NN;

\pgfplotstableread{grid time
1	0.559888
2	0.198882
3	0.177105
4	0.295043
}\NNO;

\pgfplotstableread{grid time
1	0.037685
2	0.231867
3	0.116071
4	0.416002
}\RFR;

\begin{axis}[
width=\columnwidth, height=3.5cm, 
% ybar,
% bar width=3.5pt, 
ymin=10^-4,
ymax=10^4,
ymode=log,
axis lines*=left, 
ymajorgrids, 
yminorgrids,
% xtick=data,
% xticklabels from table={\ipresults}{numProcs},
xtick={1,2,3,4},
xticklabels = {case30, case240, case1354, case1888},
ytick={0.0001,0.01, 1, 100, 10000},
xticklabel style={font=\scriptsize},
yticklabel style={font=\scriptsize},
xlabel style={font=\scriptsize},
ylabel style={font=\scriptsize},
xlabel={Electricity Grid},
ylabel={Processing Time (sec)},
clip=true,
legend style={at={(0.95,1.12)},fill=white,legend cell align=left,align=right,draw=white!85!black,font=\scriptsize,legend columns=1},
]

\addplot[mycolor0,mark=*] table[y=time] {\MATLAB};
%\addlegendentry{MATLAB}

\addplot[mycolor1,thick,dashed, mark=*] table[y=time] {\DTR}; 
%\addlegendentry{DTR}

\addplot[mycolor2,thick,dashed, mark=*] table[y=time] {\GBR}; 
%\addlegendentry{GBR}

\addplot[mycolor3,thick,dashed, mark=*] table[y=time] {\NN}; 
%\addlegendentry{NN-1}

\addplot[mycolor4,thick,dashed, mark=*] table[y=time] {\NNO}; 
%\addlegendentry{NN-2}

\addplot[mycolor5,thick,dashed, mark=*] table[y=time] {\RFR}; 
%\addlegendentry{RFR}
\end{axis}

\end{tikzpicture}
		\caption{Processing Time.}
		\label{fig:pt}
	\end{subfigure}%
	~ 
	\begin{subfigure}[t]{0.45\textwidth}
		\centering
        \begin{tikzpicture}

% COLORS 
%4-class OrRd Tokyo SIAM PP18
\definecolor{mycolor0}{HTML}{FF0000}
\definecolor{mycolor1}{HTML}{FFA500}
\definecolor{mycolor2}{HTML}{008000}
\definecolor{mycolor3}{HTML}{0000FF}
\definecolor{mycolor4}{HTML}{800080}
\definecolor{mycolor5}{HTML}{A52A2A}

% \pgfplotstableread{figures/greasy/verification/dxt.dat}\data;

% \pgfplotstableread{figures/TcAvg/lpModelIntpnt/parsed_lp_intpnt_tc_avg.dat}\ipdata;

% \pgfplotstableread{figures/greasy/greasyLpIntpnt/avg_lp_intpnt_greasy_w_binding.dat}\lpintpntWbindingAvg;

\pgfplotstableread{grid time
1	128.5956
2	152.0848
3	3688.9000
4	5654.2000
}\MATLAB;

\pgfplotstableread{grid time
1	0.108398
2	0.584414
3	03.048676
4	04.825628
}\DTR;

\pgfplotstableread{grid time
1	013.224367
2	037.471711
3	042.204298
4	0308.159138
}\GBR;

\pgfplotstableread{grid time
1	0167.893890
2	0252.848958
3	0249.343849
4	0943.171995
}\NN;

\pgfplotstableread{grid time
1	0602.285319
2	0276.297931
3	0309.123895
4	0354.222212
}\NNO;

\pgfplotstableread{grid time
1	07.313872
2	0245.455593
3	0513.059226
4	02958.223694
}\RFR;

\begin{axis}[
width=\columnwidth, height=3.5cm, 
% ybar,
% bar width=3.5pt, 
ymin=10^-4,
ymax=10^4,
ymode=log,
axis lines*=left, 
ymajorgrids, 
yminorgrids,
% xtick=data,
% xticklabels from table={\ipresults}{numProcs},
xtick={1,2,3,4},
xticklabels = {case30, case240, case1354, case1888},
ytick={0.0001,0.01, 1, 100, 10000},
xticklabel style={font=\scriptsize},
yticklabel style={font=\scriptsize},
xlabel style={font=\scriptsize},
ylabel style={font=\scriptsize},
xlabel={Electricity Grid},
ylabel={Training Time (sec)},
clip=true,
legend style={at={(0.95,0.45)},fill=white,legend cell align=left,align=right,draw=white!85!black,font=\scriptsize,legend columns=2, nodes={scale=0.65, transform shape}},
]

\addplot[mycolor0, mark=*] table[y=time] {\MATLAB};
\addlegendentry{MATPOWER}

\addplot[mycolor1,thick,dashed, mark=*] table[y=time] {\DTR}; 
\addlegendentry{DTR}

\addplot[mycolor2,thick,dashed, mark=*] table[y=time] {\GBR}; 
\addlegendentry{GBR}

\addplot[mycolor3,thick,dashed, mark=*] table[y=time] {\NN}; 
\addlegendentry{NN-1}

\addplot[mycolor4,thick,dashed, mark=*] table[y=time] {\NNO}; 
\addlegendentry{NN-2}

\addplot[mycolor5,thick,dashed, mark=*] table[y=time] {\RFR}; 
\addlegendentry{RFR}
\end{axis}

\end{tikzpicture}
		\caption{Training Time.}
		\label{fig:tt}
	\end{subfigure}
	\caption{Timing statistics for MATPOWER and various ML models. \label{fig:times}}
\end{figure*}

Processing time represents the time required to generate the response by a pre-trained ML model for 5000 instances of electricity grids representing LMP prediction. This measure reveals how long the ML model will take to process the data once the training is complete. In case of the ground truth, the processing time represents the elapsed time to solve the optimization problems of the same instances using MATPOWER and MOSEK.
Figure~\ref{fig:pt} presents a comparison of the processing time for 5000 grid snapshots using the classical optimization tools and the ML models considered in this study.

%\begin{figure}[htbp]
%    \centering
%    \includegraphics[width=0.45\textwidth]{process_times.pdf}
%    \caption{Processing Time for MATPOWER and various ML models.}
%    \label{fig:pt}
%\end{figure}

As demonstrated in Fig.~\ref{fig:pt}, MATPOWER requires up to five orders of magnitude more time than ML models to process the data. Among the ML models we assessed, DTR exhibits the lowest processing time. It is noteworthy that the processing time does not increase significantly for ML models as the grid size increases, which is not the case for MATPOWER and MOSEK.

\subsection{Training Time}
Training time is defined as the average time it takes for a ML model to learn from 5000 electricity grid snapshots. The time required for training is dependent on several factors, such as the complexity of the model, chosen hyperparameters, and the hardware used. Once trained, the model can process new data, which is usually significantly faster than the training process. Fig.~\ref{fig:tt} compares the training time of the ML models with the processing time of MATPOWER and MOSEK for 5000 instances of electricity grid data.

The comparison in Fig.~\ref{fig:tt} shows that ML models take a comparable amount of time to train as MATPOWER takes to solve 5000 instances of electricity grid. Furthermore, for larger grids, the ML models can be trained faster compared to the time it takes to solve the instances using MATPOWER. This observation is significant because as the size and complexity of the grid increases, this difference is expected to become more prominent. However, note that the training time excludes the time necessary to generate the training data. The generation of the training data is equivalent to the MATPOWER time shown in the figure.

%\begin{figure}[ht!]
%	\centering
%	\includegraphics[width=0.45\textwidth]{training_times.pdf}
%	\caption{Training Time for MATPOWER and various ML models.}
%	\label{fig:tt}
%\end{figure}

\subsection{Performance}
The performance of the models relies on the size of the training dataset. The DTR model was trained on varying dataset sizes ranging from 1000 to 50000 problem instances to see the how the MAPE changes.

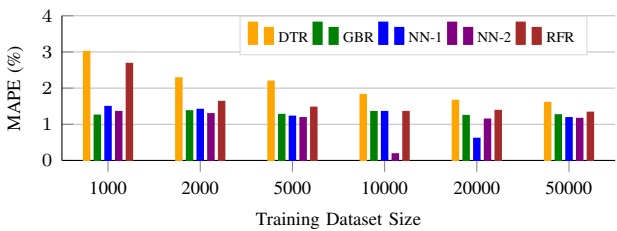
\begin{figure}[ht!]
	\centering
    \begin{tikzpicture}

% COLORS 
%4-class OrRd Tokyo SIAM PP18
\definecolor{mycolor0}{HTML}{FF0000}
\definecolor{mycolor1}{HTML}{FFA500}
\definecolor{mycolor2}{HTML}{008000}
\definecolor{mycolor3}{HTML}{0000FF}
\definecolor{mycolor4}{HTML}{800080}
\definecolor{mycolor5}{HTML}{A52A2A}

% \pgfplotstableread{figures/greasy/verification/dxt.dat}\data;

% \pgfplotstableread{figures/TcAvg/lpModelIntpnt/parsed_lp_intpnt_tc_avg.dat}\ipdata;

% \pgfplotstableread{figures/greasy/greasyLpIntpnt/avg_lp_intpnt_greasy_w_binding.dat}\lpintpntWbindingAvg;

\pgfplotstableread{grid timeDTR timeGBR timeNN1 timeNN2 timeRFR
%1	11.34  2.3 5.74 9.9 11.34
%2	5.28 1.3 2.26 1.68 11.06
3	3.00 1.24 1.48 1.34 2.67
4	2.27 1.36 1.4 1.28 1.62
5   2.18 1.26 1.21 1.17 1.46
6   1.81 1.34 1.34 0.17 1.34
7   1.65 1.23 0.6 1.13 1.37
8   1.59 1.25 1.17 1.15 1.32
}\errors;

\begin{axis}[
width=\columnwidth, height=3.5cm, 
ybar,
bar width=2pt, 
ymin=0,
ymax=4,
axis lines*=left, 
ymajorgrids, 
yminorgrids,
% xtick=data,
% xticklabels from table={\ipresults}{numProcs},
xtick={3,4,5,6,7,8},
xticklabels = {1000, 2000, 5000, 10000, 20000, 50000},
ytick={0,1,2,3,4},
xticklabel style={font=\scriptsize},
yticklabel style={font=\scriptsize},
xlabel style={font=\scriptsize},
ylabel style={font=\scriptsize},
xlabel={Training Dataset Size},
ylabel={MAPE (\%)},
clip=true,
legend style={at={(0.95,1)},fill=white,legend cell align=left,align=right,draw=white!85!black,font=\scriptsize,legend columns=5, nodes={scale=0.8, transform shape}},
]

\addplot[mycolor1, fill=mycolor1, thick] table[y=timeDTR] {\errors}; 
\addlegendentry{DTR}

\addplot[mycolor2, fill=mycolor2, thick] table[y=timeGBR] {\errors}; 
\addlegendentry{GBR}

\addplot[mycolor3, fill=mycolor3, thick] table[y=timeNN1] {\errors}; 
\addlegendentry{NN-1}

\addplot[mycolor4, fill=mycolor4, thick] table[y=timeNN2] {\errors}; 
\addlegendentry{NN-2}

\addplot[mycolor5, fill=mycolor5, thick] table[y=timeRFR] {\errors}; 
\addlegendentry{RFR}

\end{axis}

\end{tikzpicture}
	\caption{Accuracy of all models on test case 1 for \texttt{case240}.}
	\label{fig:240}
\end{figure}

As can be seen in Fig. \ref{fig:240}, the model's MAPE error decreases as the number of training samples increases for test case 1. %The rate at which the reduction happens is significant initially but slows down as the size reaches 5000 data points. 
The difference in performance between 5000 and 50000 points amounts to only a minor improvement of 0.02\% (GBR) to 0.6\% (DTR) based on the model. The idea of increasing training data points follows the law of diminishing returns. The performance gains may not justify the increased computational effort. Thus, the data set size of 5000 instances is considered in the following experiments to balance computational costs with accuracy.

Next, the performance of ML models are compared in terms of the accuracy of their predictions on the various testing cases. The error is evaluated using the MAPE \% for all the grids for different cases from tab ~\ref{tab:4.4}.

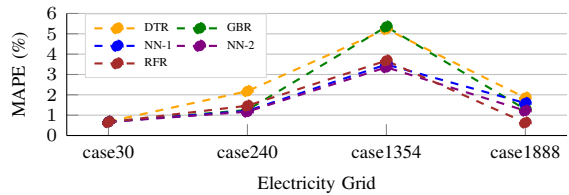
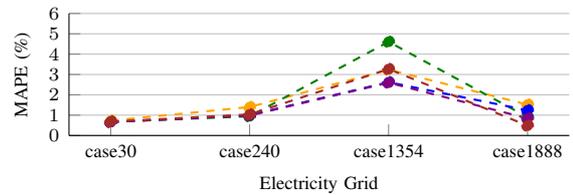
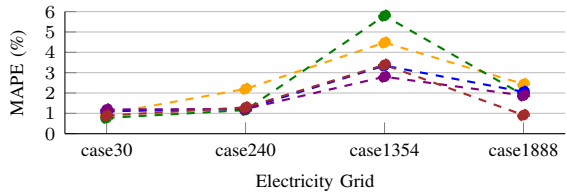
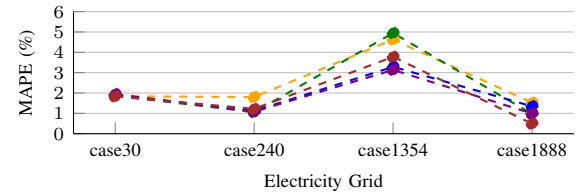
\begin{figure*}[t!]
	\centering
	\begin{subfigure}[t]{0.45\textwidth}
	    \centering
        \begin{tikzpicture}

% COLORS 
%4-class OrRd Tokyo SIAM PP18
\definecolor{mycolor0}{HTML}{FF0000}
\definecolor{mycolor1}{HTML}{FFA500}
\definecolor{mycolor2}{HTML}{008000}
\definecolor{mycolor3}{HTML}{0000FF}
\definecolor{mycolor4}{HTML}{800080}
\definecolor{mycolor5}{HTML}{A52A2A}

% \pgfplotstableread{figures/greasy/verification/dxt.dat}\data;

% \pgfplotstableread{figures/TcAvg/lpModelIntpnt/parsed_lp_intpnt_tc_avg.dat}\ipdata;

% \pgfplotstableread{figures/greasy/greasyLpIntpnt/avg_lp_intpnt_greasy_w_binding.dat}\lpintpntWbindingAvg;

\pgfplotstableread{grid time
1	00.686611
2	02.179257
3	05.230960
4	01.857539
}\DTR;

\pgfplotstableread{grid time
1 00.666235
2 01.256172
3 05.349988
4 01.288623
}\GBR;

\pgfplotstableread{grid time
1 00.673880
2 01.210856
3 03.492228
4 01.593154
}\NN;

\pgfplotstableread{grid time
1 00.665971
2 01.168791
3 03.353469
4 01.219659
}\NNO;

\pgfplotstableread{grid time
1 00.659023
2 01.463983
3 03.679439
4 00.632743
}\RFR;

\begin{axis}[
width=\columnwidth, height=3.2cm, 
% ybar,
% bar width=3.5pt, 
ymin=0,
ymax=6,
axis lines*=left, 
ymajorgrids, 
yminorgrids,
% xtick=data,
% xticklabels from table={\ipresults}{numProcs},
xtick={1,2,3,4},
xticklabels = {case30, case240, case1354, case1888},
ytick={0,1,2,3,4,5,6},
xticklabel style={font=\scriptsize},
yticklabel style={font=\scriptsize},
xlabel style={font=\scriptsize},
ylabel style={font=\scriptsize},
xlabel={Electricity Grid},
ylabel={MAPE (\%)},
clip=true,
legend style={at={(0.4,1)},fill=white,legend cell align=left,align=right,draw=white!85!black,font=\scriptsize,legend columns=2, nodes={scale=0.65, transform shape}},
]

\addplot[mycolor1,thick,dashed, mark=*] table[y=time] {\DTR}; 
\addlegendentry{DTR}

\addplot[mycolor2,thick,dashed, mark=*] table[y=time] {\GBR}; 
\addlegendentry{GBR}

\addplot[mycolor3,thick,dashed, mark=*] table[y=time] {\NN}; 
\addlegendentry{NN-1}

\addplot[mycolor4,thick,dashed, mark=*] table[y=time] {\NNO}; 
\addlegendentry{NN-2}

\addplot[mycolor5,thick,dashed, mark=*] table[y=time] {\RFR}; 
\addlegendentry{RFR}
\end{axis}

\end{tikzpicture}
		\caption{Performance of model on test case 1.}
		\label{fig:tc1}
	\end{subfigure}%
	~ 
	\begin{subfigure}[t]{0.45\textwidth}
  		\centering
        \begin{tikzpicture}

% COLORS 
%4-class OrRd Tokyo SIAM PP18
\definecolor{mycolor0}{HTML}{FF0000}
\definecolor{mycolor1}{HTML}{FFA500}
\definecolor{mycolor2}{HTML}{008000}
\definecolor{mycolor3}{HTML}{0000FF}
\definecolor{mycolor4}{HTML}{800080}
\definecolor{mycolor5}{HTML}{A52A2A}

% \pgfplotstableread{figures/greasy/verification/dxt.dat}\data;

% \pgfplotstableread{figures/TcAvg/lpModelIntpnt/parsed_lp_intpnt_tc_avg.dat}\ipdata;

% \pgfplotstableread{figures/greasy/greasyLpIntpnt/avg_lp_intpnt_greasy_w_binding.dat}\lpintpntWbindingAvg;

\pgfplotstableread{grid time
1 00.732307
2 01.400075
3 03.236266
4 01.520062
}\DTR;

\pgfplotstableread{grid time
1 00.671874
2 00.943605
3 04.608074
4 00.903449
}\GBR;

\pgfplotstableread{grid time
1 00.661610
2 01.003595
3 02.618851
4 01.240996
}\NN;

\pgfplotstableread{grid time
1 00.661122
2 00.982260
3 02.597334
4 00.842884
}\NNO;

\pgfplotstableread{grid time
1 00.676040
2 01.036929
3 03.267348
4 00.484706
}\RFR;

\begin{axis}[
width=\columnwidth, height=3.2cm, 
% ybar,
% bar width=3.5pt, 
ymin=0,
ymax=6,
axis lines*=left, 
ymajorgrids, 
yminorgrids,
% xtick=data,
% xticklabels from table={\ipresults}{numProcs},
xtick={1,2,3,4},
xticklabels = {case30, case240, case1354, case1888},
ytick={0,1,2,3,4,5,6},
xticklabel style={font=\scriptsize},
yticklabel style={font=\scriptsize},
xlabel style={font=\scriptsize},
ylabel style={font=\scriptsize},
xlabel={Electricity Grid},
ylabel={MAPE (\%)},
clip=true,
legend style={at={(0.3,1)},fill=white,legend cell align=left,align=right,draw=white!85!black,font=\scriptsize,legend columns=1, nodes={scale=0.8, transform shape}},
]

\addplot[mycolor1,thick,dashed, mark=*] table[y=time] {\DTR}; 
%\addlegendentry{DTR}

\addplot[mycolor2,thick,dashed, mark=*] table[y=time] {\GBR}; 
%\addlegendentry{GBR}

\addplot[mycolor3,thick,dashed, mark=*] table[y=time] {\NN}; 
%\addlegendentry{NN-1}

\addplot[mycolor4,thick,dashed, mark=*] table[y=time] {\NNO}; 
%\addlegendentry{NN-2}

\addplot[mycolor5,thick,dashed, mark=*] table[y=time] {\RFR}; 
%\addlegendentry{RFR}
\end{axis}

\end{tikzpicture}
		\caption{Performance of model on test case 2.}
		\label{fig:tc2}
	\end{subfigure}
	
	\begin{subfigure}[t]{0.45\textwidth}
	    \centering
        \begin{tikzpicture}

% COLORS 
%4-class OrRd Tokyo SIAM PP18
\definecolor{mycolor0}{HTML}{FF0000}
\definecolor{mycolor1}{HTML}{FFA500}
\definecolor{mycolor2}{HTML}{008000}
\definecolor{mycolor3}{HTML}{0000FF}
\definecolor{mycolor4}{HTML}{800080}
\definecolor{mycolor5}{HTML}{A52A2A}

% \pgfplotstableread{figures/greasy/verification/dxt.dat}\data;

% \pgfplotstableread{figures/TcAvg/lpModelIntpnt/parsed_lp_intpnt_tc_avg.dat}\ipdata;

% \pgfplotstableread{figures/greasy/greasyLpIntpnt/avg_lp_intpnt_greasy_w_binding.dat}\lpintpntWbindingAvg;

\pgfplotstableread{grid time
1 00.973090
2 02.200801
3 04.477876
4 02.439648
}\DTR;

\pgfplotstableread{grid time
1 00.774356
2 01.161433
3 05.797827
4 01.898453
}\GBR;

\pgfplotstableread{grid time
1 01.101975
2 01.222147
3 03.345737
4 02.058067
}\NN;

\pgfplotstableread{grid time
1 01.190868
2 01.212937
3 02.797960
4 01.876577
}\NNO;

\pgfplotstableread{grid time
1 00.890876
2 01.292798
3 03.385801
4 00.913646
}\RFR;

\begin{axis}[
width=\columnwidth, height=3.2cm, 
% ybar,
% bar width=3.5pt, 
ymin=0,
ymax=6,
axis lines*=left, 
ymajorgrids, 
yminorgrids,
% xtick=data,
% xticklabels from table={\ipresults}{numProcs},
xtick={1,2,3,4},
xticklabels = {case30, case240, case1354, case1888},
ytick={0,1,2,3,4,5,6},
xticklabel style={font=\scriptsize},
yticklabel style={font=\scriptsize},
xlabel style={font=\scriptsize},
ylabel style={font=\scriptsize},
xlabel={Electricity Grid},
ylabel={MAPE (\%)},
clip=true,
legend style={at={(0.3,1)},fill=white,legend cell align=left,align=right,draw=white!85!black,font=\scriptsize,legend columns=1, nodes={scale=0.8, transform shape}},
]

\addplot[mycolor1,thick,dashed, mark=*] table[y=time] {\DTR}; 
%\addlegendentry{DTR}

\addplot[mycolor2,thick,dashed, mark=*] table[y=time] {\GBR}; 
%\addlegendentry{GBR}

\addplot[mycolor3,thick,dashed, mark=*] table[y=time] {\NN}; 
%\addlegendentry{NN-1}

\addplot[mycolor4,thick,dashed, mark=*] table[y=time] {\NNO}; 
%\addlegendentry{NN-2}

\addplot[mycolor5,thick,dashed, mark=*] table[y=time] {\RFR}; 
%\addlegendentry{RFR}
\end{axis}

\end{tikzpicture}
		\caption{Performance of model on test case 3.}
		\label{fig:tc3}
	\end{subfigure}
	~
	\begin{subfigure}[t]{0.45\textwidth}
		\centering
        \begin{tikzpicture}

% COLORS 
%4-class OrRd Tokyo SIAM PP18
\definecolor{mycolor0}{HTML}{FF0000}
\definecolor{mycolor1}{HTML}{FFA500}
\definecolor{mycolor2}{HTML}{008000}
\definecolor{mycolor3}{HTML}{0000FF}
\definecolor{mycolor4}{HTML}{800080}
\definecolor{mycolor5}{HTML}{A52A2A}

% \pgfplotstableread{figures/greasy/verification/dxt.dat}\data;

% \pgfplotstableread{figures/TcAvg/lpModelIntpnt/parsed_lp_intpnt_tc_avg.dat}\ipdata;

% \pgfplotstableread{figures/greasy/greasyLpIntpnt/avg_lp_intpnt_greasy_w_binding.dat}\lpintpntWbindingAvg;

\pgfplotstableread{grid time
1   01.841217
2   01.800696
3   04.629563
4   01.522291
}\DTR;

\pgfplotstableread{grid time
1   01.869279
2   01.058195
3   04.946952
4   00.968926
}\GBR;

\pgfplotstableread{grid time
1   01.913869
2   01.115892
3   03.287056
4   01.345337
}\NN;

\pgfplotstableread{grid time
1   01.941382
2   01.079920
3   03.136161
4   00.990067
}\NNO;

\pgfplotstableread{grid time
1   01.832583
2   01.219676
3   03.778062
4   00.491812
}\RFR;

\begin{axis}[
width=\columnwidth, height=3.2cm, 
% ybar,
% bar width=3.5pt, 
ymin=0,
ymax=6,
axis lines*=left, 
ymajorgrids, 
yminorgrids,
% xtick=data,
% xticklabels from table={\ipresults}{numProcs},
xtick={1,2,3,4},
xticklabels = {case30, case240, case1354, case1888},
ytick={0,1,2,3,4,5,6},
xticklabel style={font=\scriptsize},
yticklabel style={font=\scriptsize},
xlabel style={font=\scriptsize},
ylabel style={font=\scriptsize},
xlabel={Electricity Grid},
ylabel={MAPE (\%)},
clip=true,
legend style={at={(0.3,1)},fill=white,legend cell align=left,align=right,draw=white!85!black,font=\scriptsize,legend columns=1, nodes={scale=0.8, transform shape}},
]

\addplot[mycolor1,thick,dashed, mark=*] table[y=time] {\DTR}; 
%\addlegendentry{DTR}

\addplot[mycolor2,thick,dashed, mark=*] table[y=time] {\GBR}; 
%\addlegendentry{GBR}

\addplot[mycolor3,thick,dashed, mark=*] table[y=time] {\NN}; 
%\addlegendentry{NN-1}

\addplot[mycolor4,thick,dashed, mark=*] table[y=time] {\NNO}; 
%\addlegendentry{NN-2}

\addplot[mycolor5,thick,dashed, mark=*] table[y=time] {\RFR}; 
%\addlegendentry{RFR}
\end{axis}

\end{tikzpicture}
		\caption{Performance of model on test case 4.}
		\label{fig:tc4}
	\end{subfigure}
	\caption{Accuracy of the ML models in terms of \texttt{MAPE} \% for all the grids for different test cases. \label{fig:errors}}
\end{figure*}

\subsubsection*{\textbf{Test Case 1}}

Fig.~\ref{fig:tc1} presents the performance of models on data from test case 1. For the \texttt{case30} grid, all ML models have achieved the MAPE under 1\%. The ML models show a slighter lower accuracy on larger grid with MAPE under 2\%, which is a trend that is consistent across all grids an exception of \texttt{case1354}.
The MAPE of the \texttt{case1354} grid ranges between 3.5\% for DNN and RFR, while DTR and GBR reached MAPE above 5\%. However, this might be still considered as acceptable error considering the trade-off with processing time. 

%\begin{figure}[htbp]
%    \centering
%    \includegraphics[width=0.45\textwidth]{testcase1.pdf}
%    \caption{Accuracy of \texttt{testcase1}}
%    \label{fig:tc1}
%\end{figure}

\subsubsection*{\textbf{Test Case 2}}

The results shown in Fig.~\ref{fig:tc2} indicate that reducing the maximum transmission capacity by approximately 10\% did not affect the performance of the ML models compared to their accuracy in test case 1. The base case electricity grid configurations in the PGLib-OPF definitions do not typically operate at maximum capacity. As a result, our data set includes very few instances where resources are pushed to the maximum when introducing perturbations. This could explain why similar levels of MAPE errors are observed in the two test cases.

%\begin{figure}[htbp]
%    \centering
%    \includegraphics[width=0.45\textwidth]{testcase3.pdf}
%    \caption{Accuracy of \texttt{testcase2}}
%    \label{fig:tc2}
%\end{figure}

\subsubsection*{\textbf{Test Case 3}}

Fig.~\ref{fig:tc3} shows the results for test case 3, where one transmission line in each electricity grid was removed. The number of transmission lines varies across the different grids, and we observed the highest performance drop in \texttt{case30}, which has only 41 transmission lines, compared to \texttt{case1888} with 2531 transmission lines. The drop in performance is more likely to occur in smaller grids because the difference between the base case and modified definitions may significantly alter the topology of the grid and the feasible power flows in the grid. Overall, the results suggest that the impact of shutting off one transmission line on the performance of ML models is grid-dependent.

%\begin{figure}[htbp]
%    \centering
%    \includegraphics[width=0.45\textwidth]{testcase5.pdf}
%    \caption{Accuracy of \texttt{testcase3}}
%    \label{fig:tc3}
%\end{figure}

\subsubsection*{\textbf{Test Case 4}}

The results for test case 4 are shown in Fig.~\ref{fig:tc4}. This test case further amplifies the impact observed in case 3 by disconnecting a generator from the grid. Since each grid has a limited number of generators, turning off a crucial one can significantly affect the feasible region of the power flows. As a result, the most significant performance drop is observed in the smallest grid, \texttt{case30}. However, the impact on model accuracy is negligible for other grids compared to the first test case.
%\begin{figure}[htbp]
%    \centering
%    \includegraphics[width=0.45\textwidth]{testcase6.pdf}
%    \caption{Accuracy of \texttt{testcase4}}
%    \label{fig:tc4}
%\end{figure}
\section{Conclusions} \label{sec:con}
This study extends the previous literature by investigating the feasibility and robustness of ML models in predicting LMP. Comparing the processing and training time of ML models with MATPOWER's processing time revealed that ML models are indeed feasible and offer a significantly faster alternative once the ML is trained. 
ML models provide acceptable accuracy, with MAPE within 6\% for all test cases, indicating the robustness of ML models even on data that simulates edge-case scenarios. We found that the DTR tends to have the highest MAPE error among popular ML models, while NN-1 and NN-2 outperformed their peers. However, NNs have a major disadvantage of considerably longer training time.
%
%Overall, this study demonstrates that ML models offer a feasible and robust alternative for predicting LMP values for electricity grids. 
Overall, by significantly reducing the processing and training time compared to traditional methods, ML models can help to reduce the overall cost of LMP predictions and forecasting. This is particularly important in future energy markets that will have high penetration from renewable energy sources, which can make prices more uncertain and volatile. 
%Additionally, a large-scale deployment of this new technology and enabling data sharing would only strengthen the overall functioning of the system.
These findings suggest that the adoption of ML models for LMP prediction has significant potential to improve the efficiency of energy markets in the future.
\bibliographystyle{ieeetr}
\bibliography{refs}

\end{document}